\documentclass[aps,prl,preprint,groupadress,showpacs]{revtex4-1}
\usepackage{graphicx,bm}
\usepackage{amsmath}
\newcommand{\udens}{{\,g/cm$^3$}}
\newcommand{\uev}{\,eV}
\newcommand{\ud}{\textrm{d}}
\newcommand{\gr}[1]{\textbf{#1}}

\begin{document}

\title{Evidence for out-of-equilibrium states in warm dense matter\\ probed by X-ray Thomson scattering }
\author{Jean Cl\'erouin}
\email  {jean.clerouin@cea.fr}
\affiliation{
CEA, DAM, DIF\\
91297 Arpajon, France}

\author{Gr\'egory Robert}
\affiliation{
CEA, DAM, DIF\\
91297 Arpajon, France}

\author{Philippe Arnault}
\affiliation{
CEA, DAM, DIF\\
91297 Arpajon, France}

\author{Christopher Ticknor}
\affiliation{Theoretical Division, Los Alamos National Laboratory\\
Los Alamos, New Mexico 87545, USA}
\author{Joel D. Kress}
\affiliation{Theoretical Division, Los Alamos National Laboratory\\
Los Alamos, New Mexico 87545, USA}

\author{Lee A. Collins}
\affiliation{Theoretical Division, Los Alamos National Laboratory\\
Los Alamos, New Mexico 87545, USA}

\date{\today}
\begin{abstract}
\

A recent and unexpected discrepancy between \textit{ab initio} simulations and the interpretation of a laser shock experiment on aluminum, probed by X-ray Thomson scattering (XRTS), is addressed. The ion-ion structure factor deduced from the XRTS elastic peak (ion feature) is only compatible with a strongly coupled out-of-equilibrium state. Orbital free molecular dynamics simulations with ions colder than the electrons are employed to interpret the experiment. The relevance  of decoupled temperatures for ions and electrons is discussed. The possibility that it mimics a transient, or metastable, out-of-equilibrium state after melting is also suggested.

\end{abstract}

\maketitle
High-energy \cite{KONI05} and x-ray free-electron lasers \cite{VINK12} are now able to produce matter in extreme states, such as found throughout the universe in planetary interiors \cite{BARA10}, brown dwarfs stars, and  neutron star crusts \cite{DALI09}. This high pressure ($>$ 1\:Mbar), high temperature ($>$ 1\:eV) regime, containing matter compressed up to a few times  ambient density is  also multi-ionized and of technological interest for inertial confinement fusion studies \cite{LIND04}. It is characterized by strong interactions between ions, leading to a microscopic liquid-like  structure. This regime, also referred to as warm dense matter (WDM),  challenges existing theories since no small parameter exists from which to formulate a perturbative approach. Therefore,  comparisons with experiments are particularly needed. The microscopic structure of such dense matter can only be diagnosed by x-ray Thomson scattering  techniques (XRTS \cite{GLEN09}). High-power x-rays can penetrate deeply inside the dense plasma and are scattered by electrons, reflecting their collective as well as single-particle behavior, depending on the experimental geometry (forward and backward scattering).   The frequency-resolved spectra of scattered x-rays further delineate electron from ion diagnostics (inelastic and elastic scattering), providing temperatures, densities, and ionization states. Perhaps the most novel aspect of the XRTS diagnostic is its high femtosecond time resolution, providing key insight into transient structure changes underlying physical and chemical processes \cite{ELSA14}. Such ultrashort x-ray pulses can probe structural dynamics during phase transitions or chemical reactions. This new tool may be hampered by the interpretation of XRTS, which requires theoretical models for the structural properties of the  material that are often developed for bulk plasmas in equilibrium.  The testing of  the different assumptions underlying the interpretation  of the  XRTS data becomes critical to further progress. \\

 \begin{figure}[!t]
\begin{center}
\includegraphics[scale=0.4]{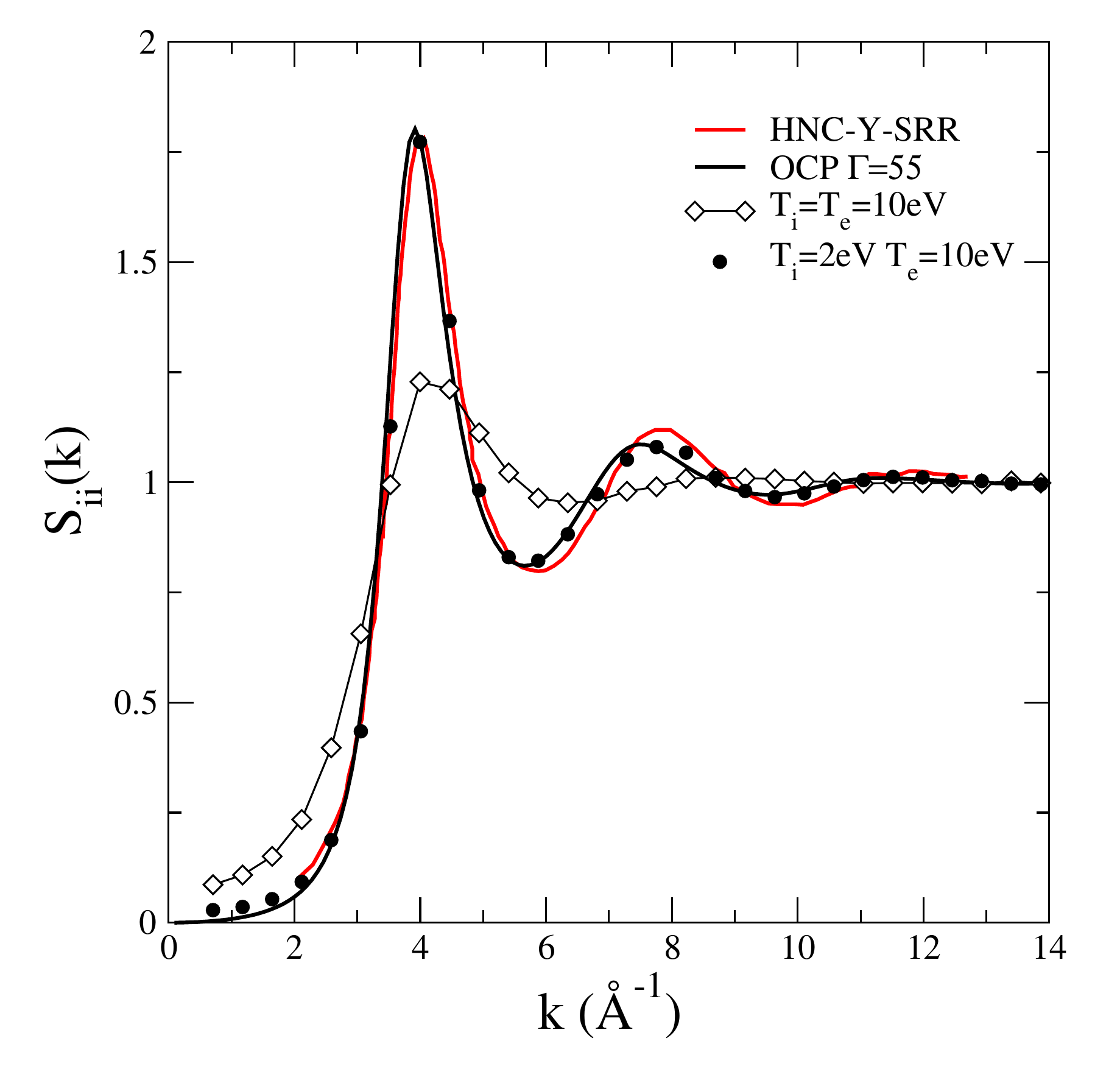}
\caption{(Color online) Ion-ion static factor computed within HNC-Y-SRR formulation and optimized to fit data \protect\cite{,MA13,MA14} (red solid) compared with the OCP result at $\Gamma$=55 (black solid). Diamonds are obtained with an equilibrium OF simulation at $T_{e}=T_{i}=10$\uev; filled circles are non-equilibrium OF simulation with $T_{i}=2$\uev, $T_{e}=10$\uev. 
\label{OCP}}
\end{center}
\end{figure} 

In this Letter, we revisit the interpretation of a recent XRTS experiment, where strong ion-ion correlations were evidenced in a laser-shocked aluminum sample \cite{MA13,MA14}. In this experiment,  in the WDM regime ($\rho \gtrsim \rho_{0}$, T $\gtrsim$ 10\uev),  the XRTS elastic peak, also called the ion feature, at different  diffusion angles, and hence at different $k$ wavevectors, shows a marked maximum at about 4\,\AA$^{-1}$, which requires a strongly-structured static ion-ion structure factor, $S_{ii}(k)$ to account for the experimental spectrum (Fig.\,\ref{OCP}). Among the different proposed theoretical approaches (Debye-H\"uckel, screened one component plasma), a hypernetted chain calculation using  a Yukawa potential plus an \textit{ad hoc}  short range repulsion (HNC-Y-SRR) best reproduces the data. From the XRTS interpretation, aluminum appears three times compressed  (8.1\udens) with an electronic  temperature of  about 10\uev\, and an average ionization $Q=3$. If the electrons and ions are in thermal equilibrium at the same temperature, $T_i=T_e$, the ion-ion coupling parameter is defined as 
\begin{equation}
  \Gamma_{ii} = \frac{Q^{2} e^2}{4\pi\epsilon_0 a_{i} k_{\rm B} T_{i}} ,
  \label{eqn:gamma}
\end{equation}
where $e$ is the fundamental charge, $a_{i} = (3/4\pi n_{i})^{1/3}$ is the mean ion sphere radius, and $n_{i}$ is the  ionic density, and  reaches a value  of 12. 
\\\indent The interpretation of this experiment has been recently extended to {\it ab initio} quantum molecular dynamics simulations in the Kohn-Sham ansatz \cite{RUTE14}. These simulations performed under equilibrium conditions (8.1\udens, 10\uev) do not reproduce the intensity of the ion feature as well as the corresponding ion structure factor.  To our knowledge, this marks the first time that  equilibrium {\it ab initio} simulations (with $T_{i}=T_{e}$)  disagree with experiments with regards to static and dynamic properties. Souza {\it et al} \cite{SOUZ14}  obtained the same ion feature, well below the experimental result, using an average atom model  with ion-ion correlations. 
 
As stressed by Ma et al \cite{MA13, MA14}, the detection of the ion-ion correlation peak at $k=4.0\,$\AA$^{-1}$ represents a new highly accurate diagnostic of the state of compression. Indeed, this peak translates into a peak in the ion structure factor $S_{ii}(k)$, which is related to the first shell of neighbors around a given ion, characteristic of a three-fold compression. At equilibrium, aluminum shocked at three-fold compression,  reaches a temperature level of 10\uev\,  on the principal Hugoniot \cite{MINA14}, in accordance with the electron temperature measurement. Nevertheless,  we investigate here the outcome of an out-of-equilibrium state defined by a more strongly-coupled ion system. As an ansatz to describe such a situation, we consider ions at a temperature $T_i$ much less than the measured electron temperature $T_e$ of 10\uev\, and present orbital free (OF) molecular dynamics simulations with  different electronic and ionic temperatures. Such simulations correspond to equilibrium states of ions and electrons separately, but to an out-of-equilibrium state of the whole system.  Since the electron system is treated  in the simulations within density functional theory at $T_e$, the ion-electron system is frozen in the out-of-equilibrium state of decoupled temperatures. \\  
\indent The OF  method, being based on a finite temperature Thomas-Fermi (TF) description of the electrons has many advantages over the orbital based (OB) molecular dynamics, which uses the Kohn-Sham ansatz. It can be easily adapted to different electronic and ionic temperatures without any  extra cost and  allows the study of large system sizes, an important consideration for computing the ion structure factor with sufficient precision. 
 Many  incursions into the hot dense regime have been done previously with OB  molecular dynamics simulations. Temperatures  as high as 500\,eV have been reached for very dense hydrogen at densities of about 160\udens ~ \cite{RECO09}, but the common limitation is about 10\,eV. These simulations are expensive,  need very hard pseudopotentials, and require a large number of orbitals in order to comply with a given level of occupancy (usually 10$^{-3}$ to 10$^{-4}$).  For electronic temperatures of 10\,eV and densities of order of 1-3 times  the normal density, the only solution to get acceptable OB simulation times is to reduce the number of atoms and hence the number of orbitals. For the conditions in Al, OB simulations were performed  with  64 ions \cite{RUTE14}. This limitation is overcome by  OF methods that can handle hundreds of ions (here 432)  for the same computational cost with only a small loss of accuracy compared to the OB results.  Details on the OF method are given in \cite{LAMB06b,LAMB13}. The transition between OB to OF simulations has been described in details in \cite{WHIT13,DANE12} and needs a full von-Weisz\"acker functional.\\
\indent     
 Here, we just recall the expression of the  finite temperature TF free energy 
$F_{0}\left[\rho\left(\gr{r}\right),R_{i}\right]$ 
\begin{eqnarray}
  \label{eq:Thomas-Fermi}
  F_{0}\left[\rho,R_{i}\right]&=&\frac{1}{\beta} \int \ud \gr{r}
\left[ \rho\left(\gr{r}\right)\Phi\left(\gr{r}\right)-\frac{2
\sqrt{2}}{3\pi^{2}\beta^{\frac{3}{2}}}I_{\frac{3}{2}}\Big(\Phi
\left(\gr{r}\right)\Big)\right] \nonumber\\
                             & &+ \int \ud \gr{r}
\,v(\gr{r},R_{i})\rho(\gr{r}) \nonumber\\
                             &&+\, \frac{1}{2} \int \ud \gr{r}\ud\gr{r'} \, \frac{\rho(\gr{r})\rho(\gr{r'})}{\vert \gr{r}-\gr{r'} \vert}+F_{xc}[\rho\gr(r)],
\end{eqnarray}
where $R_i$ stands for the nuclei positions and $\rho(\gr{r})$ is the electron density that minimizes the free energy $F_0$.  $I_{\nu}$ is the Fermi integral of order $\nu$ and $\beta$ the inverse of the electronic temperature $k_{B}T_{e}$. The total screened potential $\Phi
\left[\gr{r} \right ]$ is defined  by

\begin{equation}
\label{eq:rho}
\rho\left(\gr{r}\right)=\frac{2\sqrt{2}}{\pi^{2}\beta^{\frac{3}{2}}}
I_{\frac{1}{2}}\Big(\Phi\left(\gr{r}\right)\Big).
\end{equation}

 For the exchange-correlation term $F_{xc}$, we take the form proposed by Perrot \cite{PERR79}. The  external potential $v(\gr{r},R_{i})$ represents the Coulomb attraction of the nuclei. In OB method, electrons are separated into valence and core electrons, that are frozen in the pseudo-potential while  in OF method, the electrons all interact with the external potential. These frozen-core electrons were suspected by R\"uter and Redmer \cite{RUTE14} to be the origin of the failure of OB simulations to reproduce the data. However, we find that OF simulations at equilibrium ($T_{i}=T_{e}=10$ eV), in which all electrons interact with the potential, also do not agree with the experimental results. We emphasize that in OF simulations there is no need of pseudo-potential.  The only modification of the external potential is a  regularization at the origin to avoid the well-known  divergences of the Thomas-Fermi solution. Convergence of the results with the cutoff radius is discussed in \cite{DANE12} but is not critical here due to the rather low ionic temperature. The cutoff radius is taken here at 0.3\,a$_{i}$ and, as a check, we ran a simulation with a cutoff twice smaller without noticeable changes except on the computer time.  In the following, we present results obtained with full 3-dimensional OF molecular dynamics code for a system of  432 atoms; whereas quantities such as the atomic form factor  and regularized potentials are obtained within TF average atom model, a spherical average atom version of the OF method. \\

\indent As a guide to determine the ion temperature $T_i$ that leads to the best structure factor to reproduce the observed ion feature, we use the concept of an effective one-component plasma (OCP).  The idea is to measure the strength of the coupling of a coulombic system by a comparison of its structure (here ion structure factor) with the OCP  model \cite{HANS73}. As recalled  in Fig. 1 in Ref.\,\cite{MA14}, the OCP realizes the transition from a purely kinetic system at low $\Gamma$ coupling to a Wigner crystal at $\Gamma$ greater than  178, crossing a  strongly-coupled liquid-like regime for $20< \Gamma<120$. For liquid metals and plasmas, an effective OCP can be defined by an  adjustment of its ion structure factor with tabulated OCP structure factors. This prescription defines an effective OCP coupling parameter \cite{,OTT14}, $\Gamma_{eff}$, and hence an effective ionization state $Q$. This procedure has been successfully used in a recent study of isochorically heated dense tungsten, exhibiting the so-called $\Gamma$-plateau feature  \cite{CLER13b,ARNA13}. Nevertheless, some caution is in order since,  if it is possible to search for the best agreement between pair distribution functions, the static structure factors of real (screened systems) cannot fully agree with OCP at vanishing $k$ due to the compressibility sum-rule \cite{OCPdisp}.  With  these caveats  in mind, the comparison between the HNC-Y-SRR  structure factor of Ref.\,\cite{MA14} with the OCP one, shown in Fig.\,\ref{OCP}, suggests a coupling parameter between 50 and 60, rather than a value of 12 as indicated  in \cite{MA14} on the basis of equal ion and electron temperatures.  Fixing the aluminum valence at $Q=3$, an ionic temperature of $2$\,eV realizes this coupling. 


\begin{figure}[!t]
\begin{center}
\includegraphics[scale=0.4]{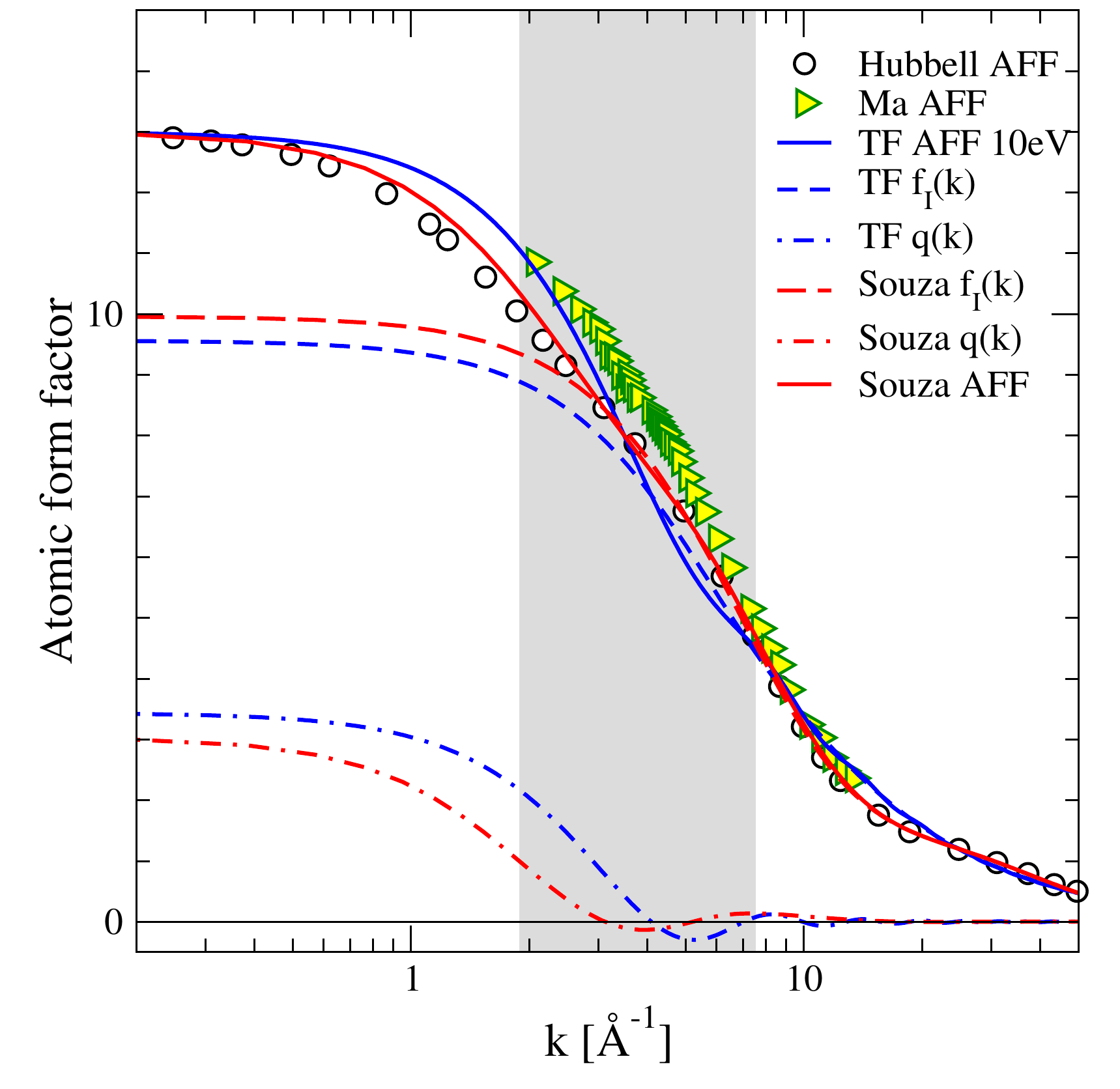}
\caption{(Color online) Atomic form factor for aluminum at 8.1\udens and $T_{e}=10$\uev. Black circles are  Hubbell's tabulated results \protect\cite{HUBB75}. (Blue) solid line is the full TF atomic form factor, the dashed line is the ion form factor $f_i(k)$ and  dot-dashed the screening term $q(k)$. Same symbols in red show Souza's results \protect\cite{SOUZ14}. Triangles are the atomic form factor deduced from Fig.2 of \cite{MA14}. The grey area shows the region where the maximum in ion feature occurs. 
\label{FF}}
\end{center}
\end{figure}

The ion structure factor  is not directly accessible in XRTS experiments but is modulated by the atomic form factor  to form the observed ion feature \cite{GLEN09}.  Following Chihara, the ion feature is approximated by \cite{CHIH00}  
\begin{equation}
\left | f_{I}(k)+q(k) \right |^{2} S_{ii}(k)\delta(\omega),
\label{ionfeat0}
\end{equation}
where $S_{ii}(k)$ is the ion structure factor. $f_{I}(k)$ is known as the ion form factor and $q(k)$ as the screening density, the sum of which defines the atomic form factor. R\"uter and Redmer \cite{RUTE14} have used Hubbell's tabulated atomic form factor for isolated atoms at zero temperature \cite{HUBB75}. Since these tables are computed at zero temperature, it seemed to us preferable to recompute the  atomic form factor within the Finite temperature TF average atom model. The total electronic density around a given nucleus is spitted into a ionic contribution $f_{I}(k)$ and a  free electron gas $q(k)$ by writing the electronic density as
\begin{equation}
\rho_{e}(r)=\left [ \rho_{e}(r)-\rho_{e}(r_{ws}) \right ]+\rho_{e}(r_{ws})
\label{split}
\end{equation}
where $\rho_{e}(r_{ws})$ is the Thomas-Fermi density at the edge of the  spherical average atom model. The atomic form factor, defined as the Fourier transform of the density
\begin{equation}
F(q,Z)=4\pi \int_{0}^{\infty} \rho_{e}(r) {\frac {\sin(qr)}{qr}}r^{2}dr,
\end{equation}
splits naturally into the two contributions  represented in Fig.\,\ref{FF}. We note that the screening term $q(k)$  has a limited range in $k$ and that the ion form factor $f_{I}(k)$  quickly merges  with the Hubbell's data. The sum of the two contributions at the origin yields the total number of electrons $Z$. The comparison with the average atom results of Souza \cite{SOUZ14} is shown in Fig.\,\ref{FF}. As expected the TF solution slightly overestimates the ionization and thus the screening component has a larger range in $k$. Nevertheless, both atomic form factors are very close to Hubbell's data. 
We have also extracted the atomic form factor  used  by Ma \textit{et al.}\,\cite{MA14} that links the ion structure factor to the ion feature. As shown in Fig.\,\ref{FF}, the Ma atomic form factor   is slightly larger than Hubbell or TF  in the region of the maximum of the experimental data.  

\begin{figure}[!t]
\begin{center}
\includegraphics[scale=0.4]{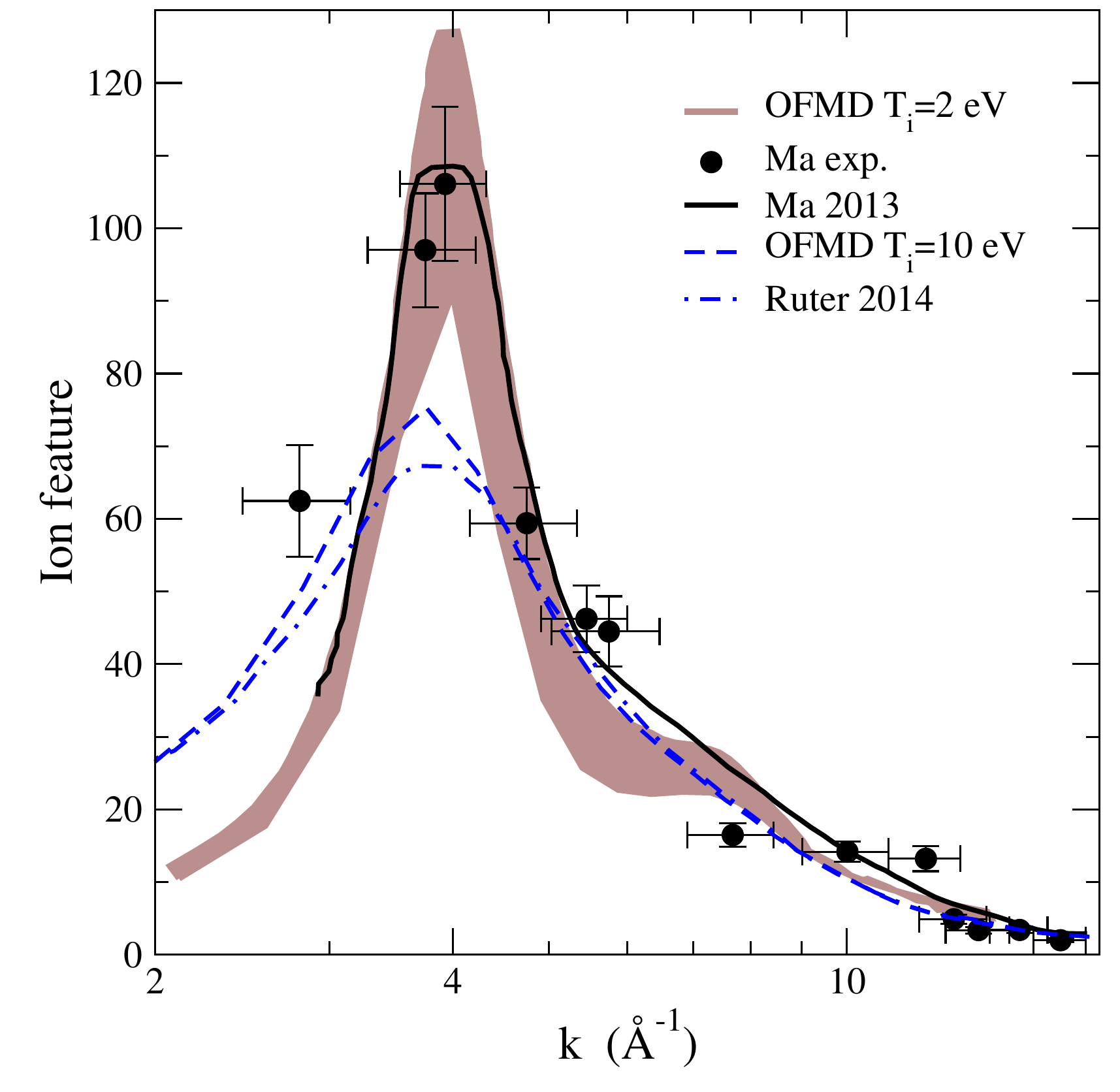}
\caption{(Color online) Circles with error bars are the experimental data of Ma et al \cite{MA13}.  Dashed line (blue)  and dot-dashed line (blue)  are the ion feature computed from an equilibrium simulation $T_{i}=T_{e}=10$\uev\,  with present OF and R\"uter's OB simulations \cite{RUTE14} respectively . The non-equilibrium OF simulation results, 8.1\udens, $T_{i}=2$\uev, and $T_{e}=10$\uev\,  are represented by the grey (brown) area  using Ma form factor (highest limit) or Thomas-Fermi atomic form factor (lowest limit). Note that the x-axis is logarithmic to emphasize the region around 4\,\AA$^{-1}$}.
\label{IF}
\end{center}
\end{figure}

Finally, the ion feature calculations are shown in Fig.\,\ref{IF}. We first show the results of our OF equilibrium simulation with $T_{e}=T_{i}=10$\uev\, at 8.1\udens. To compare with the work of R\"uter \cite{RUTE14}, we have used Hubbell's tables to compute the ion feature. Our result (dashed line) is very close to R\"uter's result (dot-dashed line), and also to Souza's result \cite{SOUZ14}. The corresponding ion structure factor  is in agreement with a coupling parameter of $\Gamma$ =12.
We now turn to the nonequilibrium OF simulations. From the effective OCP model we have chosen an ionic temperature of 2\uev\, fixing the electronic temperature at 10\uev. Because the resulting ion feature strongly  depends on the model of the atomic  form factor, we presented  in Fig.\,\ref{IF} two different models and shaded the region grey (brown) in between. The highest signal corresponds to the same atomic form factor as used by Ma {\it et al.}  in Ref.\,\cite{MA14} and the lowest one to a Thomas-Fermi atomic form factor (see Fig.\,\ref{FF}). The solid (black) curve shows the ion feature published in Fig.\,8 of Ref.\,\cite{MA14},   which results from the application of a k-vector blurring to the model, similar to those experienced in the experiment. The results are computed from our ion structure factor  shown in Fig.\,\ref{OCP}, which is close to the HNC-Y-SRR result but without any fitting  parameters.

Although our nonequilibrium OF simulations with different ion and electron temperatures reproduce the XRTS measurements, the reason why the ions could be colder than the electrons in a laser shock experiment  remains obscure. Furthermore, the temperature relaxation between ions and electrons is expected to occur on a very short time scale in this WDM regime. The relaxation rates given by the kinetic theory \cite{GERI02} are no longer valid in this regime, but the coupled-mode approaches indicate that the relaxation time scales are a factor of 100 higher than the kinetic predictions \cite{DHAR98}. This leads us to estimate that any temperature decoupling between ions and electrons should last no more than around 500\,ps. With the high temporal resolution of XRTS measurements, varying the delay between pump and probe could track the relevant relaxation process. Recently, White {\it et al.} \cite{WHIT14} used time-resolved X-ray diffraction to study electron-ion equilibration in graphite heated by fast electrons; this result indicates similar time scales. 

As an alternative, the nonequilibrium simulations with different ion and electron temperatures could actually mimic a transient, or metastable, out-of-equilibrium state after aluminum has melted. Indeed, three-fold compressed aluminum at a temperature of 10\uev\, is far from the corresponding melting temperature \cite{MINA14}. The relaxation of the ion structure factor toward its equilibrium form requires a rearrangement of the spatial configurations of ions. Such a strong nonequilibrium situation was recently observed for carbon in a similar experiment \cite{BROW14}. The time scale  for such a rearrangement can be estimated from the diffusion time $\tau = R^2/D$, with $D$ the diffusion coefficient and $R$ a characteristic correlation distance. An upper bound to this diffusion time, obtained with $R \sim 10 a_i$ and an OCP estimate of  $D$ at 1\uev\, and 8\udens\, \cite{ARNA13b}, could not exceed around 100\,ps. Again, new XRTS experiments are called for to track this phenomenon.

In summary, we conclude that the interpretation of the Ma {\it et al.}  aluminum experiment on XRTS elastic scattering \cite{MA13,MA14}  is not supported by equilibrium {\it ab initio} simulations \cite{RUTE14,SOUZ14}. We also do not agree with the suggestion of the role of core electrons  hardening interactions, since the same equilibrium calculation with an (all-electrons) OF methods leads to the same structure than a (pseudo-potential) OB one. More likely, we suspect that a nonequilibrium situation is at the heart of this discrepancy. An OF nonequilibrium calculation with T$_{i}=2$\,eV and T$_{e}=10$\,eV produces a static structure factor in excellent agreement with the one used to interpret the experiment with an arbitrary core correction. The translation of this ion structure factor into a ion feature involves the square of the atomic form factor, which produces an amount of scatter in the results depending on the theoretical model. Whether this temperature decoupling is actually at play, or ions are in a transient, or metastable, state after aluminum has melted, is still an open question. The very short timescales of relaxation in these out-of-equilibrium states should motivate new experiments with different delays between pump and probe.

\vskip 5mm
\indent T. Ma and C. Starrett are warmly acknowledged for providing their data. We also thanks F. Lambert for his essential contribution on the OFMD code. Some of the  authors (CT,JDK,LAC) gratefully acknowledge support from the Advanced Simulation and Computing Program (ASC), science campaigns 1 and 4, and LANL which is operated  by LANS, LLC for the NNSA of the U.S. DOE under Contract No. DE-AC52-06NA25396. This work was performed under the auspices of an agreement  P184 between CEA/DAM and NNSA/DP on cooperation on fundamental science.


\end{document}